\documentclass[11pt]{article}
\usepackage{moriond2004}
\usepackage{amssymb}
\usepackage{subfigure}
\usepackage{multicol}
\usepackage{graphicx}

\newcommand{\eqn}[1]
	{(#1)}

\newcommand{\tbl}[1]
	{Table~#1}

\newcommand{\fig}[1]
	{Figure~#1}
\newcommand{\Fig}[1]
	{Figure~#1}
\newcommand{\sectn}[1]
	{Section~#1}

\newcommand{\cmb}
	{{CMB}}
\newcommand{\cmbtext}
	{Cosmic Microwave Backround}

\newcommand{\cswt}
	{{CSWT}}

\newcommand{\mexhat}
	{Mexican Hat}

\newcommand{\wmap}
	{{WMAP}}

\newcommand{\astroph}
	{astro-ph}

\newcommand{\dthree}
	{{3D}}

\newcommand{\etal}
	{{\it et al.}}

\newcommand{\img}
	{\ensuremath{i}}

\newcommand{\sky}
	{\ensuremath{s}}
\newcommand{\skywav}
	{\ensuremath{S}}
\newcommand{\wav}
	{\ensuremath{\psi}}

\newcommand{\dil}
	{\ensuremath{\mathcal{D}}}
\newcommand{\scale}
	{\ensuremath{a}}
\newcommand{\effsize}
	{\ensuremath{\zeta}}
\newcommand{\rot}
	{\ensuremath{\mathcal{R}}}
\newcommand{\dmatbig}
	{\ensuremath{D}}
\newcommand{\dmatsmall}
	{\ensuremath{d}}

\newcommand{\sa}
	{\ensuremath{\omega}}
\newcommand{\saa}
	{\ensuremath{\theta}}
\newcommand{\sab}
	{\ensuremath{\phi}}
\newcommand{\eulera}
	{\ensuremath{\alpha}}
\newcommand{\eulerb}
	{\ensuremath{\beta}}
\newcommand{\eulerc}
	{\ensuremath{\gamma}}
\newcommand{\eulers}
	{\ensuremath{\eulera, \eulerb, \eulerc}}

\newcommand{\shcoeff}[1]
	{\ensuremath{\widehat{#1}}}
\newcommand{\lmax}
	{\ensuremath{l_{max}}}
\newcommand{\mmax}
	{\ensuremath{m_{max}}}
\newcommand{\conj}
	{\ensuremath{\ast}}

\newcommand{\real}
	{\ensuremath{\mathbb{R}}}
\newcommand{\sphere}
	{\ensuremath{S^2}}

\newcommand{\cswtfftterm}
	{\ensuremath{t}}

\newcommand{\ind}
	{\ensuremath{n}}
\newcommand{\num}
	{\ensuremath{N}}
\newcommand{\p}
	{\ensuremath{^\prime}}
\newcommand{\pp}
	{\ensuremath{^{\prime\prime}}}
\newcommand{\complexity}
	{\ensuremath{\mathcal{O}}}

\newcommand{\spcend}
	{\ensuremath{\:}}

\newcommand{\arcmin}
	{\ensuremath{\p}}

\begin{document}

\vspace*{4cm}
\title{A FAST DIRECTIONAL CONTINUOUS SPHERICAL\\WAVELET TRANSFORM}
\author{ J.D. MCEWEN, M.P. HOBSON, A.N. LASENBY and D.J. MORTLOCK }
\address{Cavendish Laboratory, Madingley Road, Cambridge CB3 0HE, U.K.}
\maketitle



\newlength{\statplotwidth}
\setlength{\statplotwidth}{36mm}

\newlength{\coeffmapwidth}
\setlength{\coeffmapwidth}{53mm}

\newlength{\figspacerplot}
\setlength{\figspacerplot}{5mm}

\newlength{\figspacermap}
\setlength{\figspacermap}{3mm}


\begin{abstract}

A fast algorithm for Antoine and Vandergheynst's (1998) directional Continuous
Spherical Wavelet Transform (\cswt) is presented.  
Computational requirements are reduced by a factor
of $\complexity(\sqrt{\num_{\rm pix}})$, when $\num_{\rm pix}$ is the number of
pixels on the sphere.  The spherical \mexhat\ wavelet
Gaussianity analysis of the
\wmap\ 1-year data performed by Vielva \etal\ (2003) is reproduced and
confirmed using the fast \cswt.  The proposed extension to directional
analysis is inherently afforded by the fast \cswt\ algorithm.

\end{abstract}


\section{Introduction}

A range of primordial processes imprint signatures on the temperature
fluctuations of the \cmbtext\ (\cmb).  For instance, various cosmic
defect and non-standard inflationary models predict non-Gaussian
anisotropies.
By studying the Gaussianity of the \cmb\ anisotropies 
evidence may be provided for competing scenarios of the early
Universe.  
In addition, a number of astrophysical processes introduce
secondary sources of non-Gaussianity.  Measurement
systematics or contamination may also be highlighted
by Gaussianity analysis.

Wavelets are a powerful tool for probing the Gaussianity of \cmb\
anisotropies.
Previous wavelet analysis of the \cmb, however, has been restricted
to simple spherical Haar and isotropic Mexican Hat wavelets.  A
directional analysis on the full sky has previously been prohibited by the
computational infeasibility of any implementation.  We rectify this
problem by providing a fast algorithm for Antoine and
Vandergheynst's\cite{antoine:1998} Continuous Spherical Wavelet
Transform (\cswt), based on the fast spherical convolution proposed by
Wandelt and G\'{o}rski\cite{wandelt:2001}.

The remainder of this paper is organised as follows.  The \cswt\ is
presented in \sectn{\ref{sec:cswt}} and the fast implementation in
\sectn{\ref{sec:fast}}.  
In \sectn{\ref{sec:cmb}} the fast \cswt\ is applied to reproduce the
non-Gaussianity \cmb\ analysis performed by Vielva 
\etal\cite{vielva:2003}.
Concluding remarks are made in \sectn{\ref{sec:conc}}.


\section{A directional Continuous Spherical Wavelet Transform}
\label{sec:cswt}

Antoine and
Vandergheynst\cite{antoine:1998} extend Euclidean wavelet analysis to
spherical geometry by constructing a wavelet basis on the sphere.
The natural extension of Euclidean motions on the sphere are
rotations, defined by
\mbox{$(\rot_\rho f)(\sa) = f(\rho^{-1} \sa), \, \rho \in SO(3) $,} 
where we parameterise $\rho$ by the Euler angles $(\eulers)$.
Dilations on the sphere, denoted
$(\dil_\scale f)(\sa) = f_\scale(\sa)$,
are constructed by first
lifting the sphere \sphere\ to the plane by a norm preserving
stereographic projection from the South pole,
performing the usual Euclidean dilation in the plane, before
re-projecting back onto \sphere. 
Mother spherical wavelets
are constructed simply by projecting Euclidean planar wavelets onto
the sphere by a norm preserving inverse stereographic projection.
A wavelet basis on \sphere\ may be constructed from rotations and
dilations of an admissible mother spherical wavelet.  The corresponding
wavelet family $\{ \wav_{\scale,\rho} \equiv \rot_\rho \dil_\scale \wav,
\, \rho \in SO(3), \, \scale \in \real_{\ast}^{+} \}$  provides an
over-complete set of functions in $L^2(\sphere)$.  
The \cswt\ is given by the projection
onto each wavelet basis function
\begin{equation}
\skywav(\scale, \eulers) =
\int_{\sphere}
(\rot_{\eulers} \wav_\scale)^\conj(\sa) \:
\sky(\sa) \:
d\mu(\sa)
\spcend ,
\label{eqn:cswt}
\end{equation}
where the \conj\ denotes complex conjugation and $d\mu(\sa)=\sin(\saa)
\, d\saa \, d\sab$ is the usual rotation invariant measure on the sphere.


\section{Fast algorithm}
\label{sec:fast}

A direct implementation of the \cswt\ is simply not computationally
feasible for a data set of any practical size; a fast algorithm is
essential. 
At a particular scale the \cswt\ is essentially a spherical
convolution,
hence it is possible to apply Wandelt and G\'{o}rski's\cite{wandelt:2001}
fast spherical convolution algorithm to rapidly evaluate the
transform.


\subsection{Fast implementation}
\label{sec:harmonic}

There does not exist any finite point set on the sphere that is
invariant under rotations, hence it is more natural, and in fact more
accurate for numerical purposes, to recast the \cswt\ in spherical
harmonic space.  
The Wigner rotation matrices (defined by Brink and
Satchler\cite{brink:1993}, for 
example) introduced to characterise the rotation
of a spherical harmonic may be decomposed as  
$
\dmatbig_{mm\p}^{l}(\eulers)
= e^{-\img m\eulera} \:
\dmatsmall_{mm\p}^l(\eulerb) \:
e^{-\img m\p\eulerc}
$.
This decomposition is exploited by factoring the rotation
into two separate rotations, both of which contain
a constant $\pm \pi/2$ polar rotation:
$
\rot_{\eulers}
= \rot_{\eulera-\pi/2, \; -\pi/2, \; \eulerb} \:\:
\rot_{0, \; \pi/2, \; \eulerc+\pi/2}
$.
By uniformly sampling and applying some algebra the \cswt\ may be
recast as
\begin{equation}
\skywav [\ind_\eulera, \ind_\eulerb, \ind_\eulerc] =
e^{-\img 2\pi [ 
\frac{\ind_\eulera \lmax}{\num_\eulera} +
\frac{\ind_\eulerb \lmax}{\num_\eulerb} +
\frac{\ind_\eulerc \mmax}{\num_\eulerc}]}
\sum_{j=0}^{\num_\eulera-1} \:
\sum_{j\p=0}^{\num_\eulerb-1} \:
\sum_{j\pp=0}^{\num_\eulerc-1}
\cswtfftterm_{j, j\p, j\pp} \:
e^{ \img 2\pi [
\frac{j\ind_\eulera}{\num_\eulera} +
\frac{j\p \ind_\eulerb}{\num_\eulerb} +
\frac{j\pp \ind_\eulerc}{\num_\eulerc}]}
\spcend ,
\label{eqn:cswt_fast}
\end{equation}
where the summation is simply the unnormalised \dthree\ inverse
discrete Fourier transform of 
\begin{equation}
\cswtfftterm_{m+\lmax, m\p+\lmax, m\pp+\mmax} =
e^{i(m\pp-m)\pi/2} 
\sum_{l=\max(\mid m \mid, \mid m\p \mid, \mid m\pp \mid)}^{\lmax}
\dmatsmall_{m\p m}^l(\pi/2) \:
\dmatsmall_{m\p m\pp}^l(\pi/2) \:
\shcoeff{\wav}_{lm\pp}{}^\conj \:
\shcoeff{\sky}_{lm}
\spcend ,
\label{eqn:cswt_fast_term}
\end{equation}
where $\shcoeff{\cdot}_{lm}$ denote spherical harmonic coefficients,
$\lmax$ and $\mmax$ define the general and azimuthal band limits of the
wavelet respectively and
the shifted indices show the conversion between the harmonic and
Fourier conventions. 
The \cswt\ may be performed very rapidly in spherical harmonic
space by using fast Fourier techniques to rapidly and simultaneously evaluate
\eqn{\ref{eqn:cswt_fast}}, once \eqn{\ref{eqn:cswt_fast_term}} is
precomputed.%
\footnote{Memory and computational requirements may be reduced by a
further factor of two for real signals by exploiting the conjugate
symmetry relationship
$\cswtfftterm_{-m,-m\p,-m\pp}=\cswtfftterm_{m,m\p,m\pp}^\conj$.}


\subsection{Comparison with other algorithms}

Direct and semi-fast (where only one rotation is performed in Fourier
space) implementations of the \cswt\ are also possible.  
A comparison of the theoretical complexity and typical execution times
of each algorithm is presented in \tbl{\ref{tbl:cswt_comparison}}.
The fast \cswt\ algorithm provides a saving of 
$\complexity(\sqrt{\num_{\rm pix}})$ for $\num_{\rm pix}$ pixels on
the sphere.

\begin{table}
\begin{center}
\caption{\cswt\ algorithm comparisons}
\label{tbl:cswt_comparison}
\footnotesize
\subfigure[Complexity: Let the number of
  samples of each parameter be of order $L$, except $\num_\eulerc$
  which typically may be much lower.]
{
  \begin{tabular}{ll|l|c|ll} \cline{3-4}
  &&  \textbf{Algorithm} & \textbf{Complexity} &\\ \cline{3-4}
  &&  Direct & 
  $
  \mathcal{O}(L^4
  \num_\eulerc)
  $          &&\\
  &&  Semi-fast & 
  $
  \mathcal{O}(L^3
  \log_2(L)
  \num_\eulerc)
  $           &&\\
  &&  Fast &
  $
  \mathcal{O}(L^3
  \num_\eulerc)
  $           &&\\
  \cline{3-4}
  \end{tabular}
} \quad
\subfigure[Typical execution times: Tests performed on a dual 900MHz %
  processor system with 4GB of memory.]
{
  \begin{tabular}{ll|c|ccc|ll} \cline{3-6}
  && $\mathbf{\num_{\rm pix}}$
  & \multicolumn{3}{c|}{\textbf{Execution time}} 
  \\
  &&  & \multicolumn{3}{c|}{\tiny{(min:sec)}} 
  \\
  &&  & Direct
  & Semi-fast
  & Fast
  \\ \cline{3-6}
  &&  768 & 
  00:01.19 &
  00:01.12 &
  00:00.01 &&
  \\
  &&  3,072 & 
  00:18.60 &
  00:17.38 &
  00:00.04 &&
  \\
  &&  12,288  & 
  05:01.48 &
  04:43.06 &
  00:00.21 &&
  \\
  && 786,432  & 
  - &
  - &
  01:54.15 && 
  \\
  \cline{3-6}
  \end{tabular}
}
\end{center}

\end{table}


\section{\cmb\ non-Gaussianity analysis}
\label{sec:cmb}

We reproduce the Gaussianity analysis of
Vielva \etal\cite{vielva:2003}, preprocessing the \wmap\ data in the
same manner. 
The resolution of the co-added map defined by Komatsu
\etal\cite{komatsu:2003} is down-sampled by a factor or 4, before 
the \emph{{Kp0}} exclusion mask is applied to remove emissions due to
the Galactic plane and known point sources.


\subsection{Spherical wavelet analysis}

The \cswt\ is a linear operation, hence the wavelet coefficients of a
Gaussian map will also obey a Gaussian distribution.  
To test for deviations from Gaussianity, skewness and kurtosis
statistics are calculated for each wavelet coefficient map at each scale.
Monte Carlo simulations are performed to construct confidence bounds
on the test statistics.

The application of the \emph{Kp0} mask distorts coefficients
corresponding to wavelets that overlap with the mask exclusion region.
These wavelet coefficients must be removed from any subsequent analysis.  
Our construction of an extended coefficient exclusion mask differs to
that of Vielva~\mbox{\etal\cite{vielva:2003}} and inherently
accounts for the dominant distortion (either point-source or
Galactic plane) at each scale.
The only non-zero coefficients in a \cswt\ of the original mask are
those that are distorted (due to the zero-mean property of spherical
wavelets).  These may be easily detected and the coefficient
exclusion mask extended accordingly.


\subsection{Results}

\begin{figure}
\centering
\mbox{
\subfigure[\scriptsize{Skewness}]
        {\hspace{\figspacerplot}
	 \includegraphics[width=\statplotwidth,angle=-90,clip=]
	 {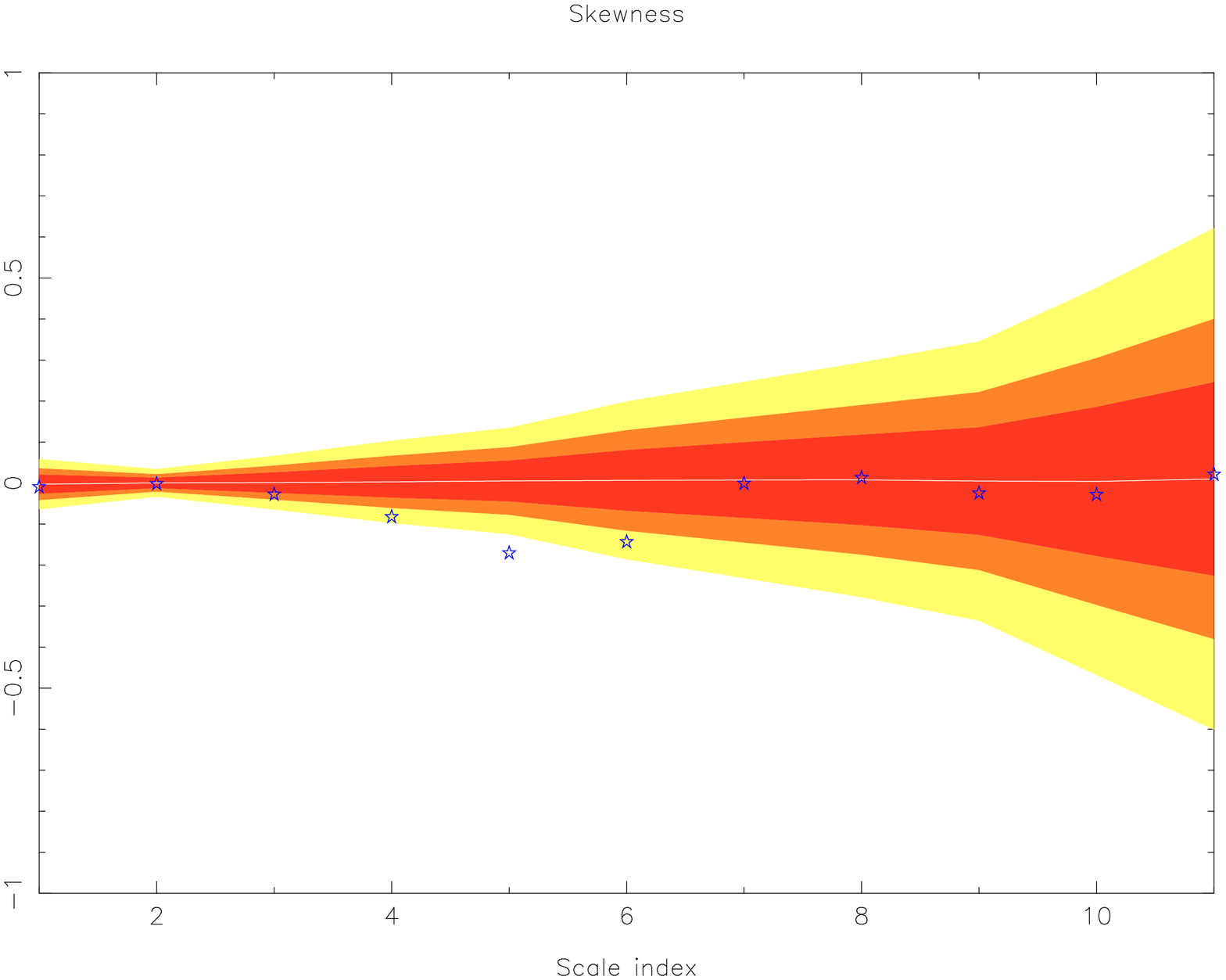} 
	 \hspace{\figspacerplot}} \quad
\subfigure[\scriptsize{Kurtosis}]
        {\hspace{\figspacerplot}
	 \includegraphics[width=\statplotwidth,angle=-90,clip=]
	 {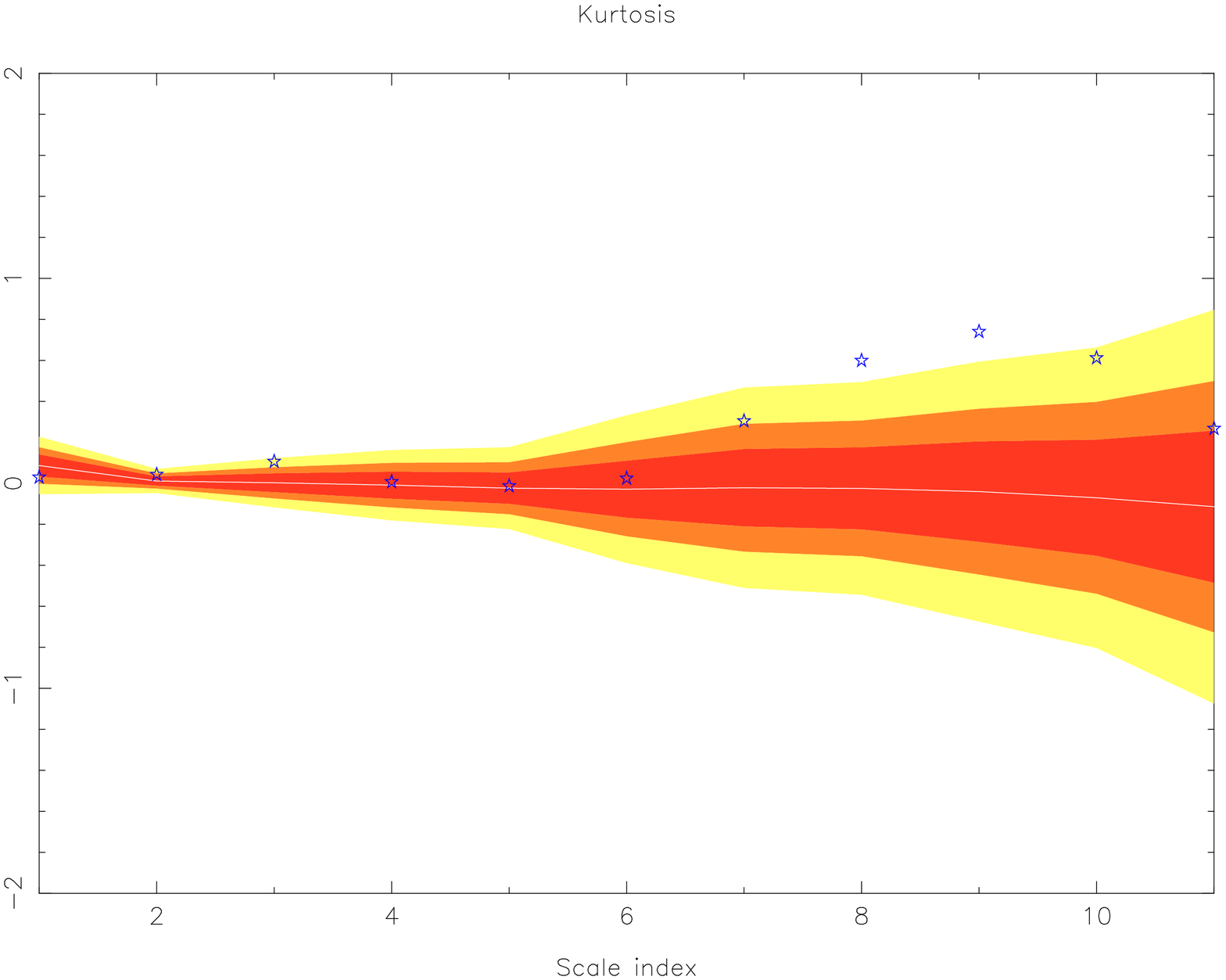} 
	 \hspace{\figspacerplot}}
}
\caption[Spherical \mexhat\ wavelet coefficient statistics]
	{Spherical \mexhat\ wavelet coefficient statistics: Confidence regions
	derived from Monte Carlo simulations are shown for 68\% (red),
	95\%
	(orange) and 99\% (yellow) levels, as is the mean (solid white line).}
\label{fig:mexhat_stats_mask}
\end{figure}

\begin{figure}
\centering
\mbox{
\subfigure[\scriptsize{Original coefficient map}]
	{\hspace{\figspacermap}
	 \includegraphics[width=\coeffmapwidth]
	 {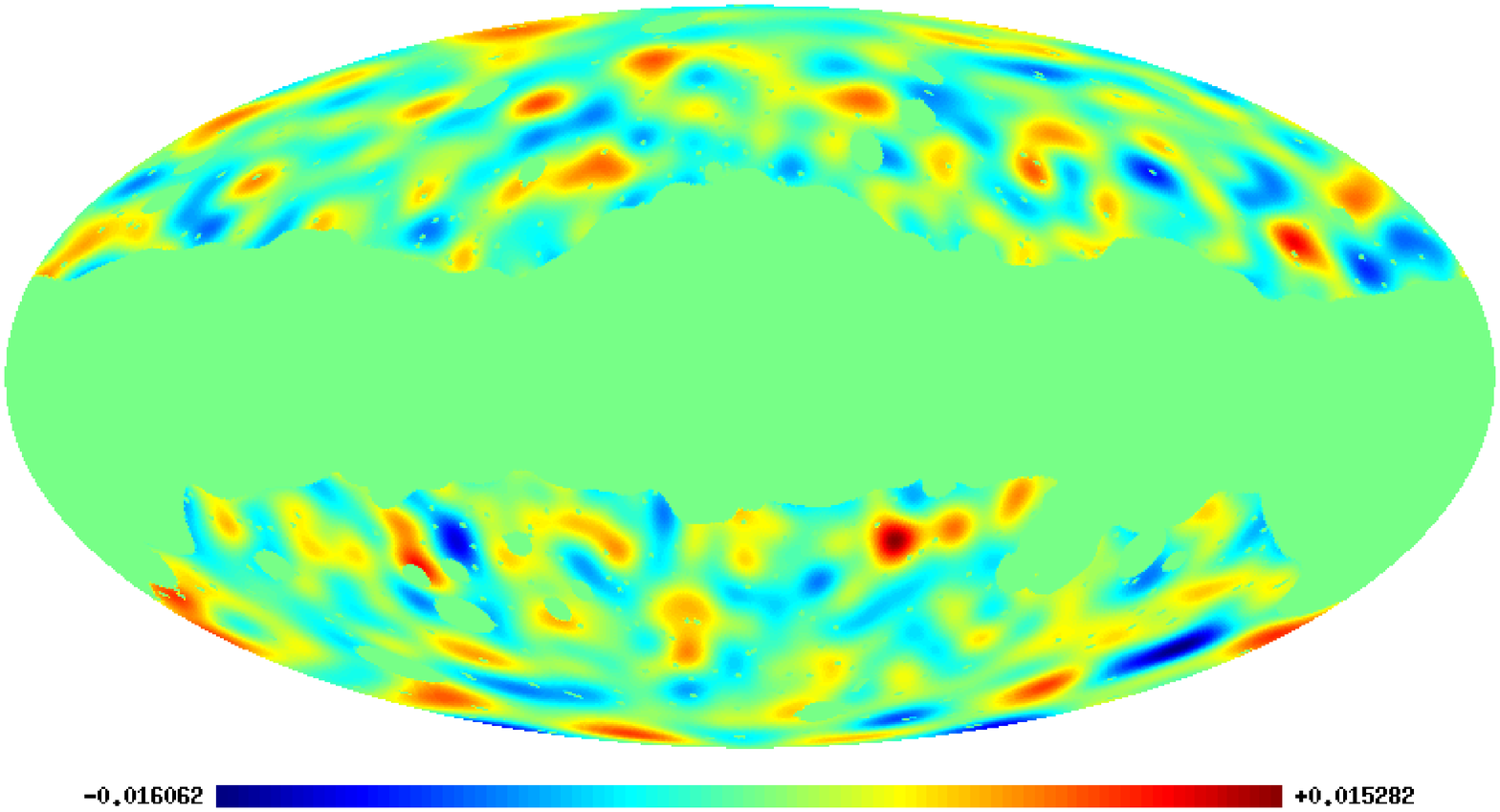}
	 \hspace{\figspacermap} } \quad
\subfigure[\scriptsize{Thresholded coefficient map}]
	{\hspace{\figspacermap} 
	 \includegraphics[width=\coeffmapwidth]
	 {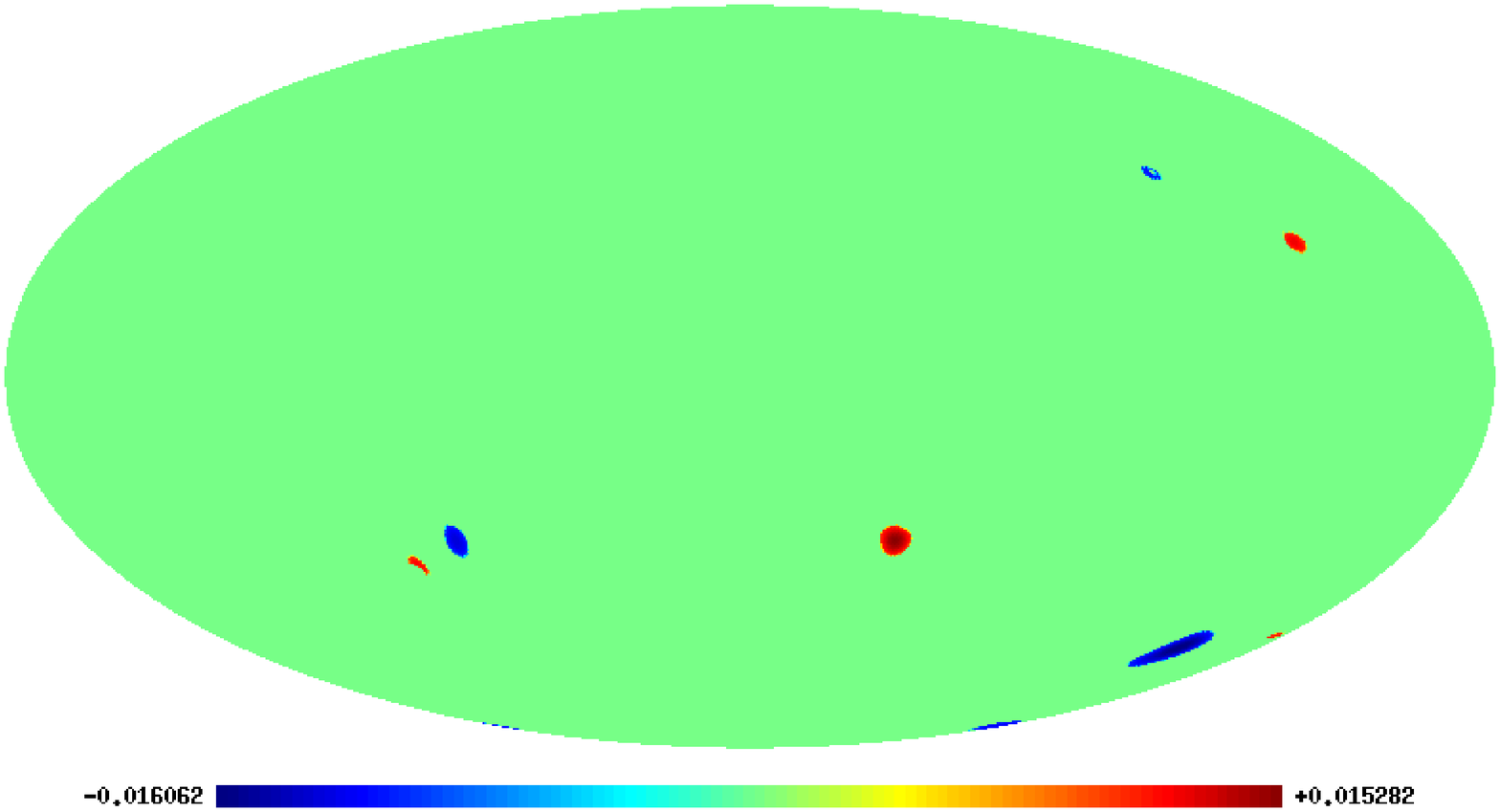}
 	 \hspace{\figspacermap}}
}
\caption[Spherical \mexhat\ wavelet coefficients at scale
	$\scale_8=250\arcmin$]
	{Spherical \mexhat\ wavelet coefficients at scale
	$\scale_8=250\arcmin$: Those coefficients below \linebreak
	$3 \, \sigma(\scale_8)$ are thresholded to zero so that
	likely deviations from Gaussianity may be localised on the
	\cmb\ sky.}
\label{fig:mexhat_coeff}
\end{figure}

We reproduce the results of Vielva \etal\cite{vielva:2003} for the spherical
\mexhat\ wavelet analysis of the co-added \wmap\ data.  The
\mexhat\ wavelet scales $\{ \scale_i \}_{i=1}^{11} = \{ 14, 
25, 50, 75, 100, 150, 200, \linebreak 250, 300, 400, 500\}_{i=1}^{11}$
acrmin are considered, corresponding to an effective size of the sky of 
\linebreak
$\effsize_i=4\tan^{-1}(  a_i/\sqrt{2} ) \approx 2\sqrt{2} \, \scale_i $
(defined as the angular separation between opposite zero-crossings).
\Fig{\ref{fig:mexhat_stats_mask}} shows the skewness and kurtosis of
the coefficients at each scale.
The wavelet analysis inherently allows one to localise signal
components on the sky, as illustrated in \fig{\ref{fig:mexhat_coeff}}.
We make similar observations to Vielva \etal\cite{vielva:2003},
although the different coefficient exclusion masks {pro\-duce} slight
discrepancies.  These discrepancies do not alter the general findings
of the analysis. 

\section{Conclusions and future work}
\label{sec:conc}

A fast algorithm is presented and evaluated for performing a
directional \cswt\ on the sphere.
The fast implementation reduces the complexity of the \cswt\ by
$\complexity(\sqrt{\num_{\rm pix}})$, where $\num_{\rm pix}$ is the number of
pixels on the sphere.  
Furthermore, the numerical accuracy of the \cswt\ is improved by
elegantly representing rotations in harmonic space.

The Gaussianity analysis of the \wmap\ 1-year data performed by Vielva
\etal\cite{vielva:2003} has been reproduced and confirmed using the
fast \cswt.  
We consider the  extension to a full directional analysis in an
upcoming publication by McEwen
\etal\cite{mcewen:2004}; preliminary findings indicate
deviations from Gaussianity outside of the 99\% confidence level.



\end{document}